\documentclass{article}[12pt]
\usepackage{graphicx}
\title{Lagrangian Crumpling Equations}
\author{Mark A. Peterson
\\Mount Holyoke College}
\begin{document}
\maketitle
\begin{abstract}
A concise method for following the evolving geometry of a moving surface using Lagrangian coordinates
is described.
All computations can be done in the fixed geometry of the initial surface despite the evolving complexity
of the moving surface.  The method is applied to three problems in nonlinear elasticity:  the bulging
of a thin plate under pressure (the original motivation for F\"oppl-von Karman theory), the
buckling of a spherical shell under pressure, and the phenomenon of capillary wrinkles
induced by surface tension
in a thin film.  In this last problem the inclusion of a gravitational potential energy term
in the total energy improves the agreement with experiment.
\end{abstract}
\section{Introduction}
The elasticity theory of thin shells is largely differential geometry by another name.
In this paper I describe a method for following the differential geometric data of
a surface as it moves, and illustrate its application to non-linear elasticity theory.  The equations of the
method are completely general for smooth surfaces, and so could in principle describe
the complex motions of crumpling up to the formation of singularities.

Problems involving elastic membranes have been approached in several ways,
including numerical simulation by triangulated surfaces, using a polyhedral approximation to
differential geometry \cite{Gompper}.  Another approach has been to use differential geometry and scaling laws
to understand the line and point singularities of crumpled surfaces analytically \cite{lobkovsky,Lobkovsky,Pomeau,Pauchard,Cerda,Witten}, and numerically\cite{DiDonna}.  The method of this paper generalizes
familiar methods of mechanical engineering for the non-linear elasticity
theory of thin shells \cite{Foppl,flugge,Niordson} in going beyond second order,
and in treating initially curved surfaces in a unified way.

Section \ref{GeomMethods} establishes notation for the differential geometry of a moving surface
and shows how to use Lagrangian coordinates to simplify its
description, the main
idea of this paper.
Section \ref{EvolutionEquations} summarizes the observations of the previous
section in a system of differential equations for the evolving surface
and its strains.  Section \ref{FvK} compares
this approach to F\"oppl-von Karman (FvK) theory, and solves the motivating
problem for that theory, the bulging of a thin rectangular plate subject to pressure, by
integrating the evolution equations forward in time.
Section \ref{Buckling} uses second order expansions of the crumpling equations to
describe the buckling of a sphere under pressure.  Section \ref{Wrinkles} uses the insights of
Cerda and Mahadevan \cite{Cerda2} to give a more detailed description of a phenomenon recently
discussed in \cite{Huang}, capillary wrinkles induced by surface tension in a thin film.  A
previously unnoticed discrepancy with experiment
is partially resolved with the inclusion of the gravitational potential
energy of the system.

\section{Geometrical Methods}
\label{GeomMethods}
In terms of smooth coordinates $(x^1,x^2,x^3)$ in space one can describe the deformation of a material object
by the trajectories of its constituent particles, solutions of equations of motion
\begin{equation}
\label{EqOfMotion}
\frac{dx^i}{dt}=V^i(x,t)
\end{equation}
where
\begin{equation}
V=V^i\partial_i
\end{equation}
is the vector field generating the flow, and $t$ is a parameter along the flow.  Integrating the
system forward to $t=1$, one can also think of $V^i$ as a displacement, a slight abuse of notation
that should be clear from context.
Metric relations among the particles are given by
\begin{equation}
\label{ds2}
ds^2=g_{ij}dx^idx^j
\end{equation}
where $g_{ij}$ is a Riemannian metric tensor, perhaps, but not necessarily, the Euclidean metric.

I coordinatize the material object by Lagrangian coordinates,
convected by the flow, i.e., every material point keeps the same coordinates that it had
originally.  In this case the changing metric relationship of material points, namely the change in
the expression Eq.~(\ref{ds2}), is due entirely to the change in the metric components $g_{ij}$, because
$dx^i$, which for this purpose simply assigns to a line segment the coordinate difference of its endpoints,
is invariant.  The rate of change as a consequence of this deformation in the components of the metric $g$, or of any second rank tensor $G$,
expressed in convected coordinates, is given by the Lie derivative\cite{schutz}\cite{frankel}
\begin{equation}
\pounds_VG(\partial_j,\partial_k)=VG_{jk}+G([\partial_j,V],\partial_k)+G(\partial_j,[\partial_k,V])
\end{equation}
Here $[~,~]$ is the Lie bracket of vector fields.
It is more common to express objects like this, derivatives of tensors which are themselves tensors,
in terms of the covariant derivative with respect to the metric connection, and to employ the conventions
of raising and lowering indices with $g_{ij}$ and its matrix inverse $g^{ij}$, such that, for example
the covector with components
\begin{equation}
V_i=g_{ij}V^j
\end{equation}
has covariant derivative with respect to $x^k$ (denoted $V_{i;k}$), in terms of the ordinary partial derivative
(denoted $V_{i,k}$) given by
\begin{equation}
V_{i;k}=V_{i,k}+\Gamma^j_{ik}V_j
\end{equation}
where the coefficients of connection $\Gamma$ are
\begin{eqnarray}
\Gamma^j_{ik}=\frac{1}{2}g^{jm}\left(g_{ik,m}-g_{mi,k}-g_{km,i}\right)
\end{eqnarray}
It is straightforward to verify for any second rank tensor $G_{\mu\nu}$ that
\begin{equation}
\pounds_VG(\partial_k,\partial_\ell)=g^{ij}\left(V_jG_{kl;i}+V_{j;k}G_{i\ell}+V_{j;\ell}G_{ki}\right)
\end{equation}
In particular, if $G$ is the metric tensor $g$, which is a covariant constant, we recover
the well known result
\begin{equation}
\pounds_Vg(\partial_k,\partial_\ell)=V_{\ell ;k}+V_{k;\ell}=2U_{k\ell}
\end{equation}
where $U$ is the rate of strain tensor of the flow $V$ (or the first order strain of the
displacement $V$).
Nothing said above was specific to three dimensions,
and therefore every statement can be interpreted as referring to a surface with a Riemannian structure
if the indices take only
two values and not three.  From now on I shall use Latin indices for tensors in three-space, and
Greek indices for tensors on a surface.

Now consider a smooth material surface $M$, so thin that one may regard it as 2-dimensional, and
let $(x^2,x^3)$ be coordinates in this surface, while $x^1=z$ is displacement along the normal
to the surface, with the positive direction chosen conventionally, such that the surface $M$ is $z=0$.
Such a coordinate system exists for a neighborhood of $M$ such that
$|z|<1/C$, where $C$ is the supremum over $M$ of both principal curvatures in absolute value.
The metric tensor in these coordinates takes the form
\begin{equation}
\label{gform}
g=\left(\begin{array}{cc}
  1 & 0 \\
  0 & g_{\mu\nu}+2zh_{\mu\nu}+z^2k_{\mu\nu}
\end{array}\right)\,.
\end{equation}
The tensor $g_{\mu\nu}$, with Greek indices taking values (2,3), is
the first fundamental form of $M$, $h_{\mu\nu}$ is the second fundamental form, and $k_{\mu\nu}=h_\mu^\lambda h_{\lambda\nu}$ is the third fundamental form.
All these tensors are associated
with the surface $M$, and not with the ambient space.  They do not depend on $z$, i.e. all $z$
dependence in Eq.~(\ref{gform}) is explicit.
The plus sign on the middle term is a conventional
choice.  On a sphere,
for example, one
could take the positive direction for $z$ to be the outer normal direction, and the principal
curvatures of the sphere to be positive.

Now let a vector field $(a,V^\mu)$ be prescribed on $M$ with normal component $a(x^2,x^3)$ and tangential
components $V^\mu(x^2,x^3)$, and extend it to a neighborhood of $M$ as
\begin{equation}
\label{WoffM}
W=a\partial_z+V^\mu\partial_\mu-zG^{'\mu\nu}a_{,\mu}\partial_\nu\,,
\end{equation}
where initially the tensor $G^{'\mu\nu}=g^{\mu\nu}$.
In a short time $\Delta t$, the flow generated by the velocity field $W$
changes the metric tensor components by approximately
\begin{equation}
\label{gOffDiagonal}
\Delta g=\Delta t\pounds_{W} g
\end{equation}
The tensor $g+\Delta g$ regarded as a tensor on 3-space expresses the ambient Euclidean geometry
in Lagrangian coordinates.
If $g+\Delta g$ is restricted to the surface $z=0$ and indices (2,3),
one has the slightly altered first fundamental form of $M$
\begin{equation}
G_{\mu\nu}=(g_{\mu\nu}+\Delta g_{\mu\nu})_{|z=0}\,,
\end{equation}
expressing the non-Euclidean geometry
of the slightly altered $M$ induced by its embedding in the ambient Euclidean space.
The term linear in $z$ in Eq.~(\ref{WoffM}) was chosen to maintain the
block diagonal form of Eq.~(\ref{gform}) to first order in $z$.  Therefore,
taking the z-derivative, one has the slightly altered second fundamental form of $M$,
\begin{equation}
H_{\mu\nu}=\left[\frac{\partial}{\partial z}(g_{\mu\nu}+\Delta g_{\mu\nu})\right]_{|z=0}\,,
\end{equation}
The third fundamental form could not be computed in this way, but it is determined by $H_{\mu\nu}$,
\begin{equation}
K_{\mu\nu}=H_{\mu\lambda}H^\lambda_{~\nu}\,.
\end{equation}
I now imagine taking a sequence of such small steps, and I will
continue to denote by $G_{\mu\nu}$ and $H_{\mu\nu}$ the evolving first
and second fundamental forms giving the Riemannian structure
on $M$ induced by the embedding in Euclidean space.  I will not make use of this Riemannian
structure for computations, however.

There is another natural Riemannian structure on $M$,
namely that given by the original, undeformed first fundamental form $g_{\mu\nu}$, together with its
associated connection, etc., which I shall continue to use, being careful not to give it erroneous
interpretations.  This Riemannian structure, unlike $G_{\mu\nu}$,
has no obvious geometrical meaning on the deformed surface,
but it is still useful in a formal way.  Another possible interpretation,
deliberately suppressing the geometrical meaning of $G_{\mu\nu}$,
is to imagine a surface that is not deformed by the flow $W$ but carries tensor fields $G_{\mu\nu}$ and $H_{\mu\nu}$,
initially coinciding with $g_{\mu\nu}$ and $h_{\mu\nu}$,
that are deformed by $W$.  That these happen to be the first and second fundamental forms of an evolving
surface is forgotten.  In this picture the undeformed $g_{\mu\nu}$ has
an obvious geometrical meaning as the metric on the underlying, unchanging surface which is the arena
for the evolving $G_{\mu\nu}$ and $H_{\mu\nu}$.

In Eq.~(\ref{WoffM}) I introduced the tensor $G^{'\mu\nu}$, initially $g^{\mu\nu}$.
More generally $G^{'\mu\nu}$ is the inverse of $G_{\mu\nu}$ as a matrix. It is a tensor field on $M$,
but it is not obtained from $G_{\mu\nu}$ by
raising indices.  Raising indices is an operation accomplished by $g^{\mu\nu}$, my chosen
Riemannian structure, not by $G^{'\mu\nu}$.
The prime on $G'$ is a reminder
that it is not some version of the tensor $G$.

I have shown how $G_{\mu\nu}$ and $H_{\mu\nu}$ change, to
first order, under a deformation $(a,V^\mu)$ of $M$, assumed now always to be
extended off $M$ as in Eq~(\ref{WoffM}).
In turn, $(a,V^\mu)$ might evolve so as to reduce at each step a free energy functional
depending on $G_{\mu\nu}$ and $H_{\mu\nu}$.  In this way I will arrive at crumpling equations, a system of differential
equations for $(a,V^\mu)$, $G_{\mu\nu}$, and $H_{\mu\nu}$, describing the evolution of $M$.  Before considering the
equation for $(a,V^\mu)$, though, there is
another issue to consider.

This formulation leaves implicit what the evolving surface actually looks like, since mere knowledge
of $G_{\mu\nu}$ and $H_{\mu\nu}$ is not a convenient description of $M$.  To keep track of the positions of points
on the surface, one should integrate Eq.~(\ref{EqOfMotion})
using components of $W(0,x^\mu)=(a,V^\mu)$ with respect to fixed Cartesian coordinate
axes.  Let $X^A(x^1,x^2,x^3,t)$ be a Cartesian coordinate function in space.  It is time independent in the physical
sense, but its functional form depends on time
because the $x^i$ evolve in time.  The 1-form $dX^A=X^A_{\,\,\,,j}dx^j$
assigns the $X^A$ component $W^A$ to the vector $W$.  This 1-form
evolves in time at the rate given by the Lie derivative
\begin{eqnarray}
\label{CartesianY}
\pounds_W dX^A(\partial_i)&=&W\,dX^A(\partial_i)+dX^A([\partial_i,W])\\
&=&W^jX^A_{\,\,\,,ij}+W^j_{\,\,\,,i}X^A_{\,\,\,,j}\\
&=&(X^A_{\,\,\,,j}W^j)_{,i}
\end{eqnarray}
Thus $U^A=U^iX^A_{i}$, the $X^A$-component of any vector field $U=U^i\partial_i$ at time $t$,
can be found using $X^A_i(x^1,x^2,x^3,t)$ solving
\begin{equation}
\label{ddXA}
\frac{\partial X^A_i}{\partial t}=\frac{\partial (X^A_jW^j)}{\partial x^i}
\end{equation}
with appropriate initial conditions.
By the definition of the coordinate $x^1=z$, the Cartesian coordinate $X^A$ is an affine linear function of $z$.
It is essential therefore to expand $X^A_jW^j$ only to first order in $z$ in Eq.~(\ref{ddXA}).
To be completely explicit, $X^A_1$ is independent of $z$ and we can represent
\begin{equation}
X^A_\mu=Y^A_\mu(x^2,x^3)+zZ^A_\mu(x^2,x^3)\,.
\end{equation}
Then Eq.~(\ref{ddXA}) says
\begin{eqnarray}
\frac{\partial X^A_1}{\partial t}&=&Z^A_\mu V^\mu-Y^A_\mu a_{,\nu}G^{'\mu\nu}\\
\frac{\partial Y^A_\mu}{\partial t}&=&(X^A_1 a+X^A_\nu V^\nu)_{,\mu}\\
\frac{\partial Z^A_\mu}{\partial t}&=&(Z^A_\nu V^\nu-Y^A_\nu a_{,\lambda}G^{'\nu\lambda})_{,\mu}
\end{eqnarray}
The linear approximation I have made
in the neighborhood of $M$ obscures the fact that if $W$ were made to carry
affine normal lines to affine normal lines exactly,
as one could always require by a suitable nonlinear extension $W$ of $(a,V^\mu)$ off $M$,
then $X^A_jW^j$ would be exactly an affine linear function of $z$ without approximation.
The evolution of $M$ is the same for any extension, however, so what looks like a linear
approximation in the method is actually exact.

As a special case, I describe motion at constant velocity, i.e., $\partial W^A/\partial t=0$ for each component $A$.
Then
\begin{equation}
0=\frac{\partial\left(X^A_jW^j\right)}{\partial t}=(X^A_kW^k)_{,j}W^j+X^A_j\frac{\partial W^j}{\partial t}
\end{equation}
Thus the components of $W$ must evolve according to
\begin{equation}
\label{StraightMotion}
\frac{\partial W^k}{\partial t}=-X^k_A(X^A_{,j}W^j)_{,i}W^i
\end{equation}
Here $X^k_A$ is the inverse of $X^A_j$, considered as a matrix.  Eq.~(\ref{StraightMotion})
for straight
line motion is recognizable as
\begin{equation}
\frac{\partial W^k}{\partial t}+W^j\nabla_j W^k=0
\end{equation}
where $\nabla_k$ is the covariant derivative with respect to the metric
connection of the Euclidean metric in 3-space expressed in the evolving
Lagrangian coordinates.  I emphasize that I have chosen, however, not to use the evolving
geometry but rather the fixed initial geometry of $M$ for all computations,
a great simplification.
\section{Evolution Equations}
\label{EvolutionEquations}
By the arguments of the previous section the surface $M$ evolves according to
\begin{eqnarray}
\label{dG}
\frac{\partial G_{\kappa\lambda}}{\partial t}&=&V^\mu G_{\kappa\lambda ;\mu} + V^\mu_{\,\,\,;\kappa}G_{\mu\lambda}
+V^\mu_{\,\,\,;\lambda}G_{\mu\kappa} + 2aH_{\kappa\lambda}\\
\label{dH}
\frac{\partial H_{\kappa\lambda}}
{\partial t}&=&aK_{\kappa\lambda}-a_{,\lambda;\kappa}+\frac{1}{2}a_{,\mu}G^{'\mu\nu}\left(-G_{\kappa\lambda ;\nu}
+G_{\nu\lambda;\kappa}+G_{\nu\kappa;\lambda}\right) \nonumber\\
& &\quad\quad +V^\mu H_{\kappa\lambda;\mu}+V^\mu_{\,\,\,;\kappa}H_{\mu\lambda}+V^\mu_{\,\,\,;\lambda}H_{\mu\kappa}\\
\label{dG'}
\end{eqnarray}
Using these relations one can find how other geometric quantities change, for example the area element $\sqrt{G}\,dx^2\,dx^3$, involving the determinant
of the first fundamental form
\begin{equation}
G=G_{22}G_{33}-G_{23}G_{32}
\end{equation}
The result is
\begin{equation}
\frac{\partial \sqrt{G}}{\partial t}=(V^\mu\sqrt{G})_{,\mu}+aG^{'\mu\nu}H_{\mu\nu}\sqrt{G}
\end{equation}
Integrating one finds $\sqrt{G}$ and hence dilation strain.
The strain tensor
\begin{equation}
\label{Gstrain}
\frac{1}{2}\left(G_{\mu\nu}-g_{\mu\nu}\right)
\end{equation}
can be found by integrating Eq.~(\ref{dG}).
A natural definition for nonlinear shear strain $S_{\mu\nu}$ is
\begin{equation}
\frac{\partial S_{\mu\nu}}{\partial t}=\frac{1}{2}\left(\frac{\partial G_{\mu\nu}}{\partial t}
-\frac{1}{\sqrt{G}}\frac{\partial \sqrt{G}}{\partial t}G_{\mu\nu}\right)\,.
\end{equation}
The subtracted term removes the contribution of dilation strain.
$S_{\mu\nu}$ is not traceless, in general, beyond first order.

\section{Comparison with F\"oppl-von Karman Approach}
\label{FvK}

A simple example illustrates the use of this formalism and points out its
relationship to
F\"oppl-von Karman (FvK) theory \cite{Foppl}.  FvK considers the equilibrium state of a thin membrane
subject to external forces and boundary conditions.  Since the metric strain within a membrane is typically
small, even for large normal displacements, it makes sense to continue to use linear stress-strain relationships.
The strain may, however, be a nonlinear function of displacement,
and therefore displacement may be nonlinearly related to stress.
FvK thus produces nonlinear equations for the equilibrium shape of an elastic membrane subject
to external stress.

Historically this idea was implemented by expanding the strain tensor to first order in tangential
displacement but second order in normal displacement.  I derive the FvK strain
by solving the evolution equations to first order in
$V^\mu$ and second order in $a$, continuing to use
the notation of previous sections, with the
initial velocity vector
\begin{equation}
W^{(0)}=a\partial_z+V^{(0)\mu}\partial_\mu-zG^{'\mu\nu}a_{,\mu}\partial_\nu
\end{equation}
of Eq.~(\ref{WoffM}).  I am
using the superscript $^{(0)}$ to indicate the initial value, which is also the zeroth approximation
for an iterative solution.  Other initial values are $g_{\mu\nu}=G^{(0)}_{\mu\nu}=\delta_{\mu\nu}$ and
$h_{\mu\nu}=H^{(0)}_{\mu\nu}=0$.  I use Picard's method to generate the solution to the differential
system Eqs.~(\ref{ddXA}), (\ref{StraightMotion}, (\ref{dG}), and $(\ref{dH})$ iteratively as a power series
in $t$, taking $M$ to be
the Euclidean plane
with the usual Cartesian coordinates.  In this case there is no distinction between indices up and indices down,
and covariant derivatives are ordinary partial derivatives.
Iterating once, and ignoring quadratic terms except in $a$ gives
\begin{eqnarray}
G^{(1)}_{\kappa\lambda}&=&\delta_{\kappa\lambda}+t(V^{(0)}_{\kappa,\lambda}+V^{(0)}_{\lambda,\kappa})\\
H^{(1)}_{\kappa\lambda}&=&-ta_{,\lambda\kappa}\\
\label{V1aa}
V^{(1)}_\mu &=&taa_{,\mu}
\end{eqnarray}
Iterating a second time, still ignoring quadratic terms except in $a$, gives
\begin{equation}
G^{(2)}_{\kappa\lambda}=\delta_{\kappa\lambda}+t(V^{(0)}_{\kappa,\lambda}+V^{(0)}_{\lambda,\kappa})
-2ta_{,\kappa\lambda}+t^2a_{,\kappa}a_{,\lambda}
\end{equation}
Finally, evaluating at $t=1$, gives the FvK metric strain
\begin{equation}
\label{G2}
\frac{1}{2}(G^{(2)}_{\kappa\lambda}-\delta_{\kappa\lambda})=\frac{1}{2}(V^{(0)}_{\kappa,\lambda}+V^{(0)}_{\lambda,\kappa}
+a_{,\kappa}a_{,\lambda})-a_{,\kappa\lambda}
\end{equation}
This is the computational starting point for FvK theory.
The rest of that
theory follows from minimizing the elastic energy,
expressed as a quadratic functional of this strain
and the first order bending strain $H^{(1)}_{\mu\nu}$,
to find the equilibrium shape.

The approach of this paper is to develop the nonlinear strain as the solution to
a differential system.
From that point of view the derivation of Eq.~(\ref{G2})
is not very natural, since to obtain it one must artificially impose the condition that the
trajectories of the particles are straight lines, a condition that introduces, via Eq.~(\ref{V1aa}),
a second order correction into the strain that is necessary to obtain Eq.~(\ref{G2}).
Although one can certainly parameterize the possible final shapes of $M$ by displacement of particles
along straight lines, it is a different thing to say that particles actually move along
straight lines.  FvK theory does not claim this, and in that sense
it is not a dynamical theory.  A dynamical theory would determine the
evolution of the velocity vector $(a,V^\mu)$ by some local physical law, replacing Eq.~(\ref{StraightMotion})
in the differential system.  It would be a simpler theory, both conceptually
and computationally, in that solving it would only
require integrating a differential system forward in time.
I will do the obvious thing and choose $W$ to reduce the elastic energy at each step, seeking the
minimum.

A typical phenomenological elastic energy functional is
\begin{equation}
\label{typicalE}
E=E_d+E_s+E_c
\end{equation}
where
\begin{eqnarray}
\label{Ed}
E_d&=&\frac{\Lambda}{2}\int_M\left(\frac{\sqrt{G}}{\sqrt{g}}-1\right)^2\sqrt{g}\,dx^2dx^3\\
\label{Es}
E_s&=&\mu\int_M S^{\kappa\lambda}S_{\kappa\lambda}\sqrt{g}\,dx^2dx^3\\
\label{Ec}
E_c&=&\frac{\kappa}{2}\int_M\left(G^{'\mu\nu}H_{\mu\nu}-g^{\mu\nu}h_{\mu\nu}\right)^2 \sqrt{g}\,dx^2dx^3
\end{eqnarray}
and where $\Lambda$, $\mu$, $\kappa$ are the 2D compression modulus, shear modulus, and
bending modulus respectively.  The area element involves $\sqrt{g}$, not $\sqrt{G}$,
because the energy due to metric strain is better understood to be per unit mass, not per unit
area, and the mass is convected with the material coordinates.
The system will move, if possible, to lower its energy,
so one must compute the variation of $E$ with respect to a small normal
displacement $\delta a$ and tangential displacement $\delta V^\mu$
\begin{equation}
\label{deltaE}
\delta E = \int_M\left[\frac{\delta E}{\delta V^\mu}\delta V^\mu+\frac{\delta E}{\delta a}\delta a\right]\sqrt{g}\,dx^2dx^3
\end{equation}
The work done on $M$ in deforming it represents energy given up by some other part of the
system, so this work should be added with a minus sign to the total change in energy.  Work
done by pressure $P$ in a small normal deformation $\delta a$, for instance, is
\begin{equation}
\label{WP}
W=P\int_M \delta a \,\sqrt{G}\,dx^2dx^3
\end{equation}
where now one must use the physical area element $\sqrt{G}dx^2dx^3$ on $M$.
A small displacement in the direction opposite to this ``gradient,"
i.e.
\begin{eqnarray}
\label{adyn}
a&=&L_a\left(-\frac{\delta E}{\delta a}+P\frac{\sqrt{G}}{\sqrt{g}}\right)\\
\label{Vdyn}
V^\mu&=&L_V\left(-\frac{\delta E}{\delta V^\mu}\right)
\end{eqnarray}
will lower the energy and move the system toward a local minimum.
The linear operators $L_a$ and $L_V$ include a projection
onto the space of admissible velocity vector fields. They must define
positive semi-definite quadratic forms with respect to the inner product
given by integration over $M$.  Apart from these requirements, they will
vary with the application.  This is just the familiar notion of conjugate
gradient.  One could also think of
$L_a$ and $L_V$ together as defining a generalized mobility tensor, because it
transforms generalized force into velocity.  If one only wants to know the final
state, one could try to choose $L_a$ and $L_V$ so as to reach equilibrium in
the most efficient way.  In any case, the dynamics of the system
is not completely determined by the elastic energies, and additional physical
considerations must be added to complete the theory in a specific application.

Eqs.~(\ref{adyn}) and (\ref{Vdyn}), together with the evolution equations
of Section \ref{EvolutionEquations}, are what I mean by Lagrangian crumpling equations.
The original problem addressed by FvK theory, the bulging of a square plate fixed
on the boundary and subject
to pressure, can be solved straightforwardly in this way.  Represent all geometric data
by discretization on a square grid of points of the original square.
Spectral methods
(fast Fourier transform with anti-aliasing) make the computation efficient,
and the gradient flow converges quickly to a solution.

\section{Buckling of a sphere under pressure}
\label{Buckling}
I consider an elastic spherical shell subject to pressure $P$, described by the phenomenological energies of
Eqs.~(\ref{Ed}), (\ref{Es}), (\ref{Ec}), and (\ref{WP}).  For small enough pressure the sphere
is uniformly compressed, but as pressure increases it buckles.  I will describe the buckling
by using expansions of strain to second order in displacement where necessary, not the FvK
expansion, but the ``dynamic" one of this paper, found by solving the crumpling equations iteratively.
It turns out that the expansion must include more terms than FvK.

For a sphere of radius $R$, in terms of spherical polar coordinates $(\theta,\phi)$,
\begin{eqnarray}
g_{\mu\nu}&=&{\rm diag}(R^2,R^2\sin^2\theta)\\
h_{\mu\nu}&=&g_{\mu\nu}/R\\
k_{\mu\nu}&=&g_{\mu\nu}/R^2
\end{eqnarray}
Taking $R=1$, and regarding all quantities now as
dimensionless, the perturbed geometric quantities in a general displacement $(a,V^\mu)$
are
\begin{eqnarray}
G_{\mu\nu}&=&g_{\mu\nu}+V_{\mu;\nu}+V_{\nu;\mu}+2ag_{\mu\nu}\\
H_{\mu\nu}&=&g_{\mu\nu}+ag_{\mu\nu}-a_{,\mu;\nu}+V_{\mu;\nu}+V_{\nu;\mu}\\
\label{sqrtG}
\sqrt{G}&=&\sqrt{g}[1+(V^\mu_{~;\mu}+2a)+\frac{1}{2}(V^\mu_{~;\mu}V^\nu_{~;\nu}
+V^\mu V^\nu_{~;\nu;\mu}\nonumber \\
&+&4aV^\mu_{~;\mu}+2V^\mu a_{,\mu}-aa^\mu_{~;\mu}+2a^2)]
\end{eqnarray}
The area element $\sqrt{G}$ had to be found to second order in displacement.
To first order in displacement the shear strain in the sphere is
\begin{equation}
S_{\mu\nu}=\frac{1}{2}(V_{\mu;\nu}+V_{\nu;\mu}-V^\lambda_{~;\lambda}g_{\mu\nu})
\end{equation}
Parameterize the displacement by coefficients $(a_{\ell m},b_{\ell m}, c_{\ell m})$, such that
\begin{eqnarray}
a&=&\sum_{\ell m} a_{\ell m} Y_{\ell m}\\
V^\mu&=&g^{\mu\nu}\sum_{\ell m}b_{\ell m}Y_{\ell m,\nu}+\epsilon^{\mu\nu}\sum_{\ell m}c_{\ell m}Y_{\ell m,\nu}
\end{eqnarray}
where the $Y_{\ell m}$ are spherical harmonics
and $\epsilon_{32}=-\epsilon_{23}=\sin\theta$, $\epsilon_{22}=\epsilon_{33}=0$ is the antisymmetric tensor.
Then for example the change in the mean curvature of the perturbed sphere is
\begin{equation}
\delta H=G^{\mu\nu}H_{\mu\nu}-g^{\mu\nu}h_{\mu\nu}=\sum_{\ell m}[\ell(\ell+1)-2]a_{\ell m}Y_{\ell m}
\end{equation}
so that the curvature energy is
\begin{equation}
E_c=\frac{\kappa}{2}\sum_{\ell m}[\ell(\ell+1)-2]^2|a_{\ell m}|^2
\end{equation}
It vanishes for $\ell=1$, as it must by Galilean invariance,
and it is independent of the tangential
displacement $V^\mu$.  The other energy expressions are
\begin{eqnarray}
W&=&4\pi Pa_{00}Y_{00}-P\sum_{\ell m}\ell(\ell+1)a_{\ell m}b_{\ell m}+2P\sum_{\ell m}|a_{\ell m}|^2\nonumber \\
&+&\frac{1}{2}Pa_{00}Y_{00}\sum_{\ell m}[-2\ell(\ell+1)a_{\ell m}b_{\ell m}+\ell(\ell+1)|a_{\ell m}|^2
+2|a_{\ell m}|^2]\\
E_d&=&\frac{\Lambda}{2}\sum_{\ell m}[-\ell(\ell+1)b_{\ell m}+2a_{\ell m}]^2 \nonumber \\
&+&\Lambda a_{00}Y_{00}\sum_{\ell m}[-2\ell(\ell+1)a_{\ell m}b_{\ell m}+\ell(\ell+1)|a_{\ell m}|^2
+2|a_{\ell m}|^2]\\
E_s&=&\frac{\mu}{2}\sum_{\ell m}\ell(\ell+1)[\ell(\ell+1)-2](|b_{\ell m}|^2+|c_{\ell m}|^2)
\end{eqnarray}
These expansions have been carried out to second order in all coefficients, but they
anticipate that $a_{00}$ is the same order as $|a_{\ell m}|^2$ for $\ell>1$, so that some
terms quadratic in $a_{00}$ appear to be third order.  I also anticipate that the
first response to pressure is a uniform compression
\begin{equation}
|a_{00}|\sim \frac{P}{ \Lambda}
\end{equation}
so that consistency requires $P<<\Lambda$.
Now seek the minimum of the total energy
\begin{equation}
E_{\rm tot}=W+E_d+E_s+E_c
\end{equation}
by choice of $(a_{\ell m},b_{\ell m},c_{\ell m})$.  Ignoring corrections of order $P/\Lambda$
gives
\begin{eqnarray}
c_{\ell m}&=&0\\
b_{\ell m}&=&\frac{2\Lambda a_{\ell m}}{\Lambda\ell(\ell+1)+\mu[\ell(\ell+1)-2]}\\
\label{a00}
a_{00}&=&-\frac{\pi PY_{00}}{\Lambda}\nonumber \\
&-&\sum_{\ell m}\frac{|a_{\ell m}|^2}{4}Y_{00}[\ell(\ell+1)-2]
\left[1+\frac{4\mu}{\Lambda\ell(\ell+1)+\mu[\ell(\ell+1)-2]}\right]
\end{eqnarray}
Putting these expressions back into $E_{\rm tot}$ gives
\begin{equation}
E_{\rm  tot}=-\pi P^2/\Lambda+Q
\end{equation}
where $Q$ is a diagonal quadratic form in the
coefficients $a_{\ell m}$.  One must determine the sign of the diagonal elements in $Q$, since
the appearance of negative coefficients in $Q$
corresponds to the onset of buckling in the corresponding mode $\ell$.
Introducing the notation
\begin{equation}
\alpha=\ell(\ell+1)\,,
\end{equation}
the diagonal element
is $(\alpha-2)F(\alpha)$, where
\begin{equation}
F(\alpha)=-\frac{P}{2}+\frac{2\Lambda\mu^2(\alpha-2)+2\mu\alpha\Lambda^2}{[\Lambda\alpha+\mu(\alpha-2)]^2}
+\kappa(\alpha-2)
\end{equation}
It is clear that for any $\alpha> 2$ the diagonal element becomes negative for large enough pressure $P$,
so that buckling must occur, but the only relevant value of $\alpha$ is the one for which this first
happens as $P$ increases.  If $\kappa>(\kappa)_{\rm cr}$, where
\begin{equation}
(\kappa)_{\rm cr} = \frac{\mu R^2}{2}\left(1+\frac{\mu}{\Lambda}\right)
\end{equation}
(I have restored dimensional factors of $R$), then $F(\alpha)$ is monotonically increasing for $\alpha>2$.  Thus as $P$ increases,
$F(\alpha)$ first becomes negative for the lowest nontrivial shape mode $\ell=2$
corresponding to $\alpha=6$, and the buckling will be of quadrupole shape.
If, on the other hand, $\kappa<(\kappa)_{\rm cr}$, the more interesting case, then $F$ has a local minimum for some
$\alpha>2$, and hence a buckling mode that doesn't simply grow from the translation mode but appears at
a higher $\alpha$.
Values of $(\alpha,P)$
for which $F(\alpha)$ has a double root correspond to the onset of buckling into this mode.
Solving $F(\alpha)=0$ and $F'(\alpha)=0$
simultaneously, and restoring dimensional factors $R$,
one finds the buckling mode $\ell_b$ and buckling pressure $P_b$
\begin{eqnarray}
\label{lb}
\ell_b(\ell_b+1)&=&\frac{2\mu\kappa+\sqrt{2R^2\mu\kappa\Lambda(\Lambda+\mu)}}{(\Lambda+\mu)\kappa}
\approx R \sqrt{\frac{2\mu\Lambda}{(\Lambda+\mu)\kappa}}\\
P_bR^3&=&\frac{\sqrt{32R^2\kappa\mu\Lambda(\Lambda+\mu)}-4\kappa\Lambda}{\Lambda+\mu}
\approx 4 R \sqrt{\frac{2\mu\Lambda\kappa}{\Lambda+\mu}}
\end{eqnarray}
According to Eq.~(\ref{lb}), the wavelength $\lambda$ of the buckling mode has the form
$\lambda\sim(\kappa/\Lambda)^{1/4}$
argued by Cerda and Mahadevan \cite{Cerda2}, although the mechanism is not quite
the same as the one they describe.  In their case the applied stress is anisotropic, while here the
symmetry breaking is spontaneous.

The second order expansion does not determine how the crumpling proceeds once buckling has occurred,
but it does give an initial condition for the crumpling equations, which are now just a differential
system of equations for $(a_{\ell m}, b_{\ell m}, c_{\ell m})$.  Solving this system numerically might
be tractable, since the the right hand side of the system involves only
integrals of products of spherical harmonics and their covariant derivatives over the sphere.

In doing this problem of the buckling sphere, I noticed that second order expansions
of strains on curved surfaces must include terms not only quadratic in normal displacement $a$,
which is the FvK prescription,
but also mixed terms like $aV^\mu$.  Eq.~(\ref{sqrtG}) contains such a term, for example,
since nothing is omitted there, through second order.
If such terms are mistakenly ignored, second order expansions of elastic energies fail to be Galilean invariant
for the simple reason that in translating a curved surface normal and tangential displacements
are necessarily of the same order.  It is not true that tangential displacements are
small even when normal displacements are large, which is the FvK argument for ignoring them.
This problem with the translation mode
$(\ell=1)$  also affects nearby $\ell$'s, by continuity.  That the expansion is then only
accurate for large $\ell$ means that it is good only if the wavelength of the perturbation $Y_{\ell m}$
is much less than the radius of curvature, but this is just the case in which we can regard the
surface as flat.  That is, simply generalizing the FvK strain of Eq.~(\ref{G2}) to a curved
surface is not much of an advance over assuming the surface to be flat.

\section{Capillary Wrinkles}
\label{Wrinkles}
A recent paper described radial wrinkles produced in a floating thin film by the surface
tension of a small drop placed at the center \cite{Huang}.  Through a combination of physical
arguments, dimensional analysis, and experiment, the phenomenon was explained in a
sufficiently quantitative way to become a useful assay for the properties of the film.
The theory is that of Cerda and Mahadevan \cite{Cerda2}.  In that paper, physical intuition
simplifies the problem, which is essentially a problem of FvK theory, but at the expense of
some of the details.  The
methods of this paper, guided by the intuition of \cite{Cerda2}, stay closer to FvK theory and show in a little more detail how the result emerges.
I also incorporate a term that turns out to be important but that was not included in the original
discussion, the gravitational potential energy of the supporting fluid, disturbed by the
wrinkling film.

A thin film disk of radius $R$ floats on a water surface, subject to surface tension $\sigma$,
and a small water drop of radius $\rho$ is placed at its center.  It is equivalent to think of an
annular film subject to radial stress $\sigma$ at its outer radius $R$ and radial stress $2\sigma$ at
its inner radius $\rho$.  Choose units so that $\rho=1$, and use cylindrical polar coordinates $(z,r,\theta)$
in the same formalism as in other sections.  The equilibrium state is attained by
a displacement $(a,V^\mu)$ that minimizes the
total energy $E$, given by the sum of the elastic energies, Eqs.~(\ref{Ed}), (\ref{Es}), and (\ref{Ec}),
the negative of the work done by surface tension
\begin{equation}
W=2\sigma\int_0^{2\pi}V^2(\rho,\theta)\rho\,d\theta -\sigma\int_0^{2\pi}V^2(R,\theta) R\,d\theta\,,
\end{equation}
and the gravitational potential energy of the water
\begin{equation}
E_g=\frac{1}{2}\int_0^{2\pi}\int_\rho^R \rho_W g a^2 r\,dr\,d\theta
\end{equation}
where $\rho_W$ is the density of water.

Expand the total energy $E$ in the sense of FvK
theory, that is, use strains linear in $V^\mu$ and quadratic in $a$.  Then
in the absence of wrinkling (i.e., $a=0$), $E$ is minimized by the radial displacement
\begin{equation}
V^2=-\frac{A}{r}+Br
\end{equation}
with
\begin{equation}
A=\frac{\sigma R^2}{2\mu (R^2-1)}\,,\quad\quad\quad B=\frac{\sigma (R^2-2)}{2\Lambda (R^2-1)}
\end{equation}
in which the disk is slightly dilated and sheared.  In the process the energy is lowered by
\begin{equation}
\Delta E= -\frac{\sigma^2\pi R^2}{2\mu(R^2-1)}-\frac{\sigma^2\pi(R^2-2)^2}{2\Lambda(R^2-1)}\,.
\end{equation}
A still lower energy is attained, however (and this is a variational estimate),
by a state with $m$ radial wrinkles of the form
\begin{eqnarray}
a&=&\frac{\alpha\cos(m\theta)}{r^\beta}\\
V^2&=&-\frac{A}{r}+Br -\frac{m^2\alpha^2}{8(\beta+1)r^{2\beta+1}}\\
V^3&=& -\frac{m\alpha^2\sin(2 m \theta)}{8r^{2\beta}}
\end{eqnarray}
The form of $V^\mu$ is chosen to cancel the $m^2$ term in the shear strain due to the wrinkle $a$ (that is the idea
of \cite{Cerda2} translated into the language of this paper).  One is still free to choose the
parameters $A$, $B$, $\beta$, $m$, and $\alpha$, this last being the
dimensionless amplitude of the wrinkles.  The dependence on $A$ and $B$ is quadratic,
so that the best values are trivially found.  The energy $E$ then has the form
\begin{equation}
\label{Ealpham}
E=\sum_{i,j=0}^{2}E_{ij}(\beta)\alpha^{2i}m^{2j}
\end{equation}
with $E_{01}=E_{02}=0.$  Minimizing with respect to $\alpha^2$ leads to
\begin{equation}
\alpha^2=-\frac{E_{11}+2E_{12}m^2}{E_{21}+2E_{22}m^2}
\end{equation}
Since the coefficient $E_{12}$ in the numerator comes from the bending energy alone, it is negligible compared
to $E_{11}$, and thus in the wrinkling regime
$\alpha\sim 1/m$, in agreement with the intuition of \cite{Cerda2}.  Substituting this value back into
Eq.~(\ref{Ealpham}), one finds that the optimal $m^2$ satisfies a cubic equation
\begin{equation}
\label{cubicm2}
0=A_1m^6+A_2m^4+A_3m^2+A_4
\end{equation}

Finally one should seek the optimal value for $\beta$.
For all physically reasonable values of the parameters in the problem, the optimal
value turns out to be $\beta=0$ (only approachable as a limit) corresponding to
wrinkles that keep a constant amplitude.

The optimal values computed above turn out to be insensitive to the dilation strain, due to equal stress $\sigma$
at inner and outer radii, and only sensitive to the unbalanced stress $\sigma$ in the center
due to the drop, creating shear
strain, the only strain that can be relieved by wrinkling.
Taking the limit $\beta\rightarrow 0$, and also ignoring $\kappa/\Lambda$ and
$\kappa/\mu$, since the bending modulus is small for thin films, leads to simple values for
the coefficients in Eq.~(\ref{cubicm2}),
\begin{eqnarray}
A_1&=&-2E_{12}E_{22}\approx -\pi^2\Lambda\kappa/64\\
A_2&=&-3E_{12}E_{21}\approx -3\pi^2\mu\kappa/256\\
A_3&=&2E_{22}E_{10}-4E_{20}E_{12}-E_{11}E_{21}\approx \pi^2\rho_WgR^2\Lambda/64+\pi^2\mu\sigma/256\\
A_4&=&E_{10}E_{21}-2E_{20}E_{11}\approx \pi^2 \rho_W g R^2\mu/256
\end{eqnarray}
The roots of Eq.~(\ref{cubicm2}), for typical physical values, are determined almost entirely
by $A_1$ and $A_3$, that is,
\begin{equation}
m\approx (-A_3/A_1)^{1/4}
\end{equation}
in formal agreement with \cite{Cerda2}.  There are two regimes, depending on the relative importance
of the gravitational term in $A_3$.  If the gravitational term is unimportant,
\begin{equation}
m\approx\left(\frac{\mu}{4\Lambda}\right)^{1/4}\left(\frac{\sigma}{\kappa}\right)^{1/4}\,.
\end{equation}
Since $\mu\approx\Lambda$, the dimensionless first factor is about $1/\sqrt{2}\approx 0.7$.
This factor was measured experimentally in \cite{Huang} and found to be about 3.6.  Because of
the 4th root, the discrepancy is very large.  If the gravitational term dominates in $A_3$,
\begin{equation}
\label{mfactor}
m\approx \left(\frac{\rho_W g R^2}{\sigma}\right)^{1/4}\left(\frac{\sigma}{\kappa}\right)^{1/4}
\end{equation}
with a crossover between the two regimes at
\begin{equation}
R\approx \sqrt{\frac{\mu\sigma}{4\Lambda\rho_W g}}\approx 1.4\,\,\,{\rm mm},
\end{equation}
taking the value $\sigma=72\times 10^{-3}$~J/m$^2$ from \cite{Huang}.  Since $R$ in that experiment
was 11.4 mm, it was in the second regime, and the dimensionless first factor in Eq.~(\ref{mfactor}) is
roughly 2, still not 3.6, but closer!

Like reference \cite{Huang}, this analysis does not explaine the observed length of the wrinkles,
which seem here to have length $R$.  A solution
going beyond second order,
obtained by integrating the crumpling equations forward in time, might resolve this
question.

\section*{Acknowledgements}
I thank Leo van Hemmen for suggesting the problem of the buckling sphere and for introducing me to FvK theory.

\end{document}